\documentclass[aps,prl,twocolumn,showpacs,floatfix]{revtex4}
\usepackage{graphicx}

\newcommand{\be}{\begin{equation}}
\newcommand{\ee}{\end{equation}}
\newcommand{\bea}{\begin{eqnarray}}
\newcommand{\eea}{\end{eqnarray}}

\begin{document}

\title{Multi-Avalanche Correlations in Directed Sandpile Models}
\author{R. Rajesh}
\email{rrajesh@imsc.res.in}
\affiliation{
Institute of Mathematical Sciences, CIT Campus, Taramani, Chennai-600113,
India}

\date{\today}

\begin{abstract}
Multiple avalanches, initiated by simultaneously toppling neighbouring sites,
are studied in three different directed sandpile models. It is argued that,
while the single avalanche exponents are different for the three models, a
suitably defined two-avalanche distribution has identical exponents. 
The origin of this universality is traced to particle
conservation. 
\end{abstract}

\maketitle

The sandpile model is a paradigm for self-organized criticality wherein 
long range correlations are generated without  any parameter being fine
tuned \cite{btw1987,btw1988}. The original version of the model and its
variants (see \cite{dhar1999} for a review) have a common feature: 
slow driving during which particles are added to the system, and fast 
dissipation during which the system relaxes through avalanches. The 
steady state is characterized by power law correlations.

Conservation laws are known to constrain correlation functions of
driven--dissipative systems. A well known example is the Kolmogorov
$4/5$-th law of three dimensional fluid turbulence
\cite{K41,K42,Grisha2,frischBook}. The conserved quantity is energy
which is pumped in at large length scales and dissipated through
viscosity at small length scales. The $4/5$-th law states that, in the
inertial range [distances $r$ between driving and dissipation length
scales], 
$\langle [v_l(\vec{r},t) - v_l(\vec{0},t)]^3 \rangle = 
- \frac{4}{5} \epsilon r$, 
where $v_l(\vec{r},t)$ is the longitudinal component of the velocity at
point $\vec{r}$ at time $t$, and 
$\epsilon$ is the energy dissipation rate. The linear dependence on $r$ 
remains true in all dimensions, while the proportionality constant is a 
function of dimension. 
There are other examples, mainly from turbulence, of a correlation function being
determined by the constant flux of a conserved quantity. Examples include
magneto-hydrodynamics \cite{pouquet2000}, burgers turbulence \cite{Grisha2} 
and advection of a passive scalar (see \cite{Grisha1} and references within).
These relations are central to understanding turbulence, acting as checkpoints
for phenomological theories. Examples outside turbulence are few.
In a recent paper \cite{CRZcfrshort}, this relation was generalised
to an arbitrary driven dissipative system that showed the general
features of turbulence.
Exact results were obtained for specific models, namely
wave turbulence \cite{zakharovBook} and models of diffusing--aggregating 
particles \cite{CRZcfrlong,CRZinout}. 

In sandpile models, the total number of particles is conserved in each
toppling. As a consequence, can any correlation function be determined?
In this paper, we
answer this question in the context of directed sandpile models. Consider
multiple avalanches obtained by adding particles simultaneously 
at nearby lattice sites. It is argued that a suitably defined
two avalanche joint probability distribution function [defined
later] will play the role of the three point velocity correlations in
the Kolmogorov $4/5$-th law, and will have a scaling exponent which is independent
of dimension and hence identical to the mean field answer.

\begin{figure}
\includegraphics[width=65mm]{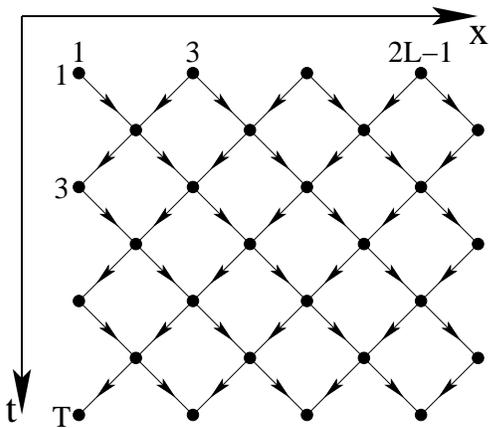}
\caption{\label{fig1}
The directed square lattice with $L$ sites along the $x$-axis
and $T$ sites along the $t$-axis. Periodic boundary conditions are
applied in the horizontal direction.
}
\end{figure}

We define the three sandpile models studied in this paper on a directed
square lattice of horizontal extent $L$ and vertical extent $T$ 
(see Fig.~\ref{fig1}). Periodic boundary conditions are imposed in
the $x$-direction and open boundary conditions along the $t$-direction, also
referred to as the time direction.
The number of particles at a site $(x,t)$ is denoted by a  non-negative 
integer $h(x,t)$. All the three models are driven by adding a particle to a 
randomly chosen site on the top layer ($t=0$) and then letting the system 
relax according to the following rules of evolution.
\\
{\it The deterministic model} \cite{dhar-ramaswamy1989}: 
A stable configuration has all $h(x,t)=0,1$.
If $(h,x,t)\geq 2$, then it relaxes by transferring two particles,
one each to its two downward neighbours, {\it i.e}, $h(x,t)$ 
decreases by $2$ and 
$h(x-1,t+1)$ and $h(x+1,t+1)$ increase by one. \\
{\it The stochastic model} \cite{paczuski-bassler2000,kloster2001}: 
This model has the same rules of evolution as the 
deterministic model except for one difference. The toppling is now
stochastic. When a site $(x,t)$ topples, 
with probability $1/4$ both
particles go to $(x-1,t+1)$, with probability $1/4$ both particles go to
$(x+1,t+1)$, with probability $1/2$, $(x-1,t+1)$ and $(x+1,t+1)$ receive one
particle each.\\
{\it The sticky model} \cite{tadic-dhar1997}: In this model, the heights can take any non-negative
integer value. A site is considered unstable if $h(x,t)\geq 2$ and it 
received at least one particle the previous time step. All unstable sites relax
simultaneously as follows: With probability $p$, the height decreases by $2$
and a particle is added to each of its downward neighbours. With probability
$(1-p)$, the site becomes stable without losing any particles.

In all the three models, if a site at the bottom ($t=T$) topples,
then the height at that site reduces by two, and the two particles are
removed from the system. An avalanche is defined as the number of topplings
that the system undergoes after a particle is added to a stable
configuration. In the steady state, the probability of an avalanche of size
$s$ is a power law distribution $P(s,T) \sim s^{-\tau} f(s T^{-\delta})$,
when $L \gg T^{1/z}$, where $z$ is the dynamic exponent. These two exponents
are not independent from each other. Particle conservation from layer to
layer results in the scaling relation $\langle s \rangle \sim T$ (for
example, see \cite{dhar1999}), implying
that $\delta(2-\tau)=1$. 

The three models belong to three different universality classes. For the
deterministic model, first studied in Ref.~\cite{dhar-ramaswamy1989},
$\tau=4/3$, $\delta=3/2$ in $d=2$, $\tau=3/2$, $\delta=2$ in $d>3$. In $d=3$,
the mean field results have logarithmic corrections \cite{dhar-ramaswamy1989}.
Stochasticity in the toppling rules is known to change the universality class
of sandpile models \cite{manna1991}. For the stochastic model, it was argued
that $\tau=10/7$, $\delta=7/4$ in $d=2$, $\tau=3/2$, $\delta=2$ in $d>3$,
with $d=3$ having the mean field exponents with logarithmic corrections 
\cite{paczuski-bassler2000,kloster2001}. The sticky model was
introduced in Ref.~\cite{tadic-dhar1997}. Introducing stickiness changes the
universality class of the sandpile model away from deterministic and
stochastic classes \cite{mohanty-dhar2002}.  The avalanche exponents are then
related to
the exponents of directed percolation. From the best numerical estimates for
directed percolation exponents, it was shown that $\tau \approx 1.32$,
$\delta=1.47$ in $d=2$ \cite{tadic-dhar1997}. In addition to having different
exponents, the sticky model is not abelian, unlike the other two models.
\begin{figure}
\includegraphics[width=\columnwidth]{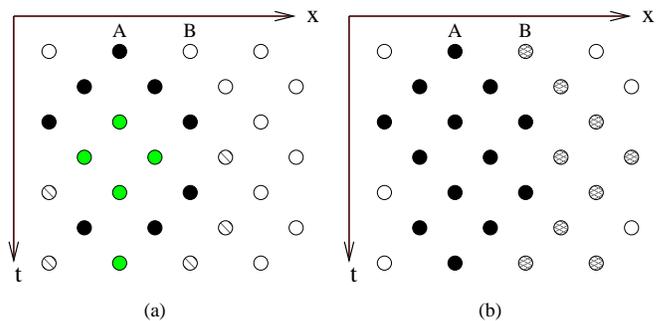}
\caption{\label{fig2}
(a) The avalanche when site $A$ is toppled is shown by filled circles (black
or grey). The black circles have height $0$ after the avalanche. The grey
circles have the same height as before the avalanche. The circles with
diameter drawn are not part of the avalanche, but will have height $1$ after
the avalanche.
(b) The avalanche when $B$ is toppled after $A$ is shown by hatched circles. 
The left boundary of the avalanche $B$ is adjacent to the right boundary 
of avalanche $A$. 
}
\end{figure}

We now define the two-avalanche distributions. Consider avalanches
initiated by adding two particles simultaneously at nearby lattice sites
(denoted by 1 and 2) on the top level. Let the set of sites belonging to
the avalanche associated with site 1 (site 2) be denoted by $S_1$ ($S_2)$.
When a site topples, if it had received
particles from only sites in $S_1$ ($S_2)$, then the site
is assigned to $S_1$ ($S_2)$. 
If on the other hand, it had received particles from sites belonging
to $S_1$ as well as $S_2$, then the site is assigned randomly to one of the
sets. Let $s_1$ ($s_2$) denote the number of topplings undergone by sites in
$S_1$ ($S_2)$. We will denote the joint probability distribution by 
$P_1(s_1,s_2)$.
For abelian models, we can also define a two avalanche distribution as
follows. Topple a site. Let the avalanche size be $s_1$. Then topple the
neighbouring site. Let the avalanche distribution be $s_2$. Let the joint
probability distribution be denoted by $P_2(s_1,s_2)$.
In this paper, it is argued that $P_1$ and $P_2$ have scaling exponents $3$ 
in all dimensions, {\it i.e.},
\be
P_i(\Lambda s_1, \Lambda s_2) = \frac{1}{\Lambda^3} P_i(s_1,s_2), \quad
i=1,2.
\label{eq:sandpilecfr0}
\ee

We give a heuristic argument
supporting this conjecture.
Consider the single avalanche probability $P(s,t)$. 
In continuous time, it schematically obeys the
equation
\be
\frac{d P(s,t)}{dt} \sim \int ds_1 ds_2 P_1(s_1,s_2) \delta(s_1+s_2-s),
\label{eq:sandpilecfr1}
\ee
where $P_1$ is the two avalanche distribution defined above. Use the fact
that, for all the three models  $\langle s \rangle
\simeq J t$, where $J$ is a constant
\cite{dhar1999}. 
Multiply Eq.~(\ref{eq:sandpilecfr1}) by $s$ and integrate over $s$. The
left hand side is a constant independent of $t$ and $s$.
Equation~(\ref{eq:sandpilecfr1}) then reduces to
\be
{\rm const} \sim \int ds ds_1 ds_2 s P_1(s_1,s_2) \delta(s_1+s_2-s).
\label{eq:sandpilecfr2}
\ee
A dimensional analysis of the right hand side of Eq.~(\ref{eq:sandpilecfr2})
immediately predicts
\be
P_1(s,s) \sim \frac{1}{s^3},
\ee
or more generally Eq.~(\ref{eq:sandpilecfr0}) with $i=1$. 
For the abelian versions of
the model, the order of toppling is not crucial. Hence, one can conjecture 
that instead of simultaneous toppling, the toppling could be sequential
and that $P_2$ obeys the same scaling law as in Eq~(\ref{eq:sandpilecfr0}).

We now give a direct proof that, in the deterministic model, 
$P_2(s_1,s_2)$ has the scaling as in Eq.~(\ref{eq:sandpilecfr0}). 
In the steady state,
each configuration of the model has equal weight \cite{dhar-ramaswamy1989}.
Thus, each height is $0$ and $1$ with probability $1/2$ independent of other sites.
The avalanches then have no holes {\it i.e.}, an avalanche is described by
the two boundaries, each of which are random walkers that annihilate
on contact. 
The clusters also have the following property. Consider the right boundary.
If it is at $(x,t)$, the next time step, it can either go to $(x+1,t+1)$ or
$(x-1,t+1)$. If it goes to $(x+1,t+1)$, then $h(x,t)=0$. It it goes to
$(x-1,t+1)$, then $h(x+1,t+1)=1$. Similar rules exist for the left walker.
An example is shown in Fig.~\ref{fig2}(a) with site $A$ having been toppled. 
Now consider the case when $B$ (see Fig.~\ref{fig2}(b)) is
toppled. The black circles cannot topple because they have height $0$, 
ensuring that the two avalanches do not overlap.
On the other hand, the sites with height $1$ will
necessarily topple provided the avalanche survives up to that level. Hence the
right boundary of first avalanche and the left boundary of second avalanche
will be adjacent to each other (see Fig.~\ref{fig2}(b)).

The calculation of $P_2(s_1,s_2)$ now reduces to the problem of three
annihilating  walkers. Let us calculate the probability that both avalanches
exceed time $t$. This is equal to the survival probability of three 
annihilating random walkers up to time $t$, which varies as $t^{-3/2}$ when
$t \gg 1$ \cite{FG1988}. 
Using the scaling $s\sim t^{3/2}$ \cite{dhar-ramaswamy1989}, we
obtain that $\int_s^\infty \int_s^\infty ds_1 ds_2 P_2(s_1,s_2) \sim s^{-1}$, 
or $P_2(s,s) \sim s^{-3}$, consistent
with Eq.~(\ref{eq:sandpilecfr0}). The argument for $P_1(s_1,s_2)$ proceeds
on exactly the same lines and we omit the argument here.
\begin{figure}
\includegraphics[width=\columnwidth]{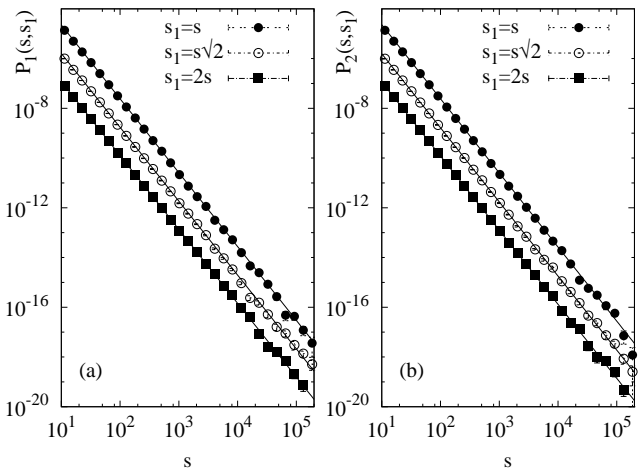}
\caption{\label{fig3}
The variation of (a)$P_1(s,s_1)$ and (b) $P_2(s,s_1)$ with $s$ is shown 
for the deterministic model for
$s_1=s,s\sqrt{2},s$. The bottom two curves have been shifted for clarity.
The solid lines are power laws with exponent (a) $2.97$ and (b) $2.98$.
}
\end{figure}
\begin{figure}
\includegraphics[width=\columnwidth]{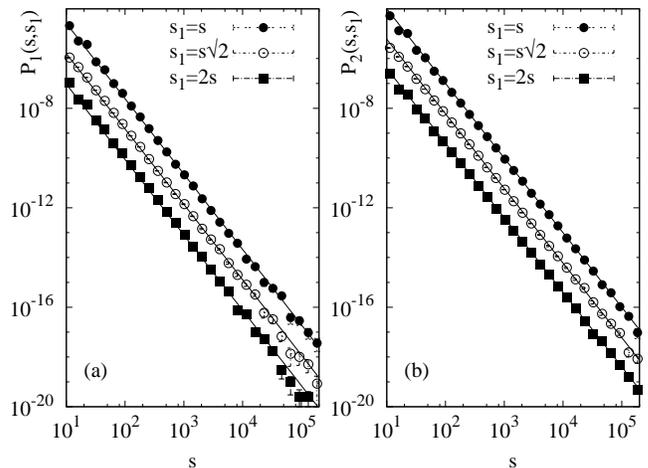}
\caption{\label{fig4}
The variation of (a)$P_1(s,s_1)$ and (b) $P_2(s,s_1)$ with $s$ is shown 
for the stochastic model for
$s_1=s,s\sqrt{2},s$. The bottom two curves have been shifted for clarity.
The solid lines are power laws with exponent (a) $3.03$ and (b) $3.00$.
}
\end{figure}
\begin{figure}
\includegraphics[width=\columnwidth]{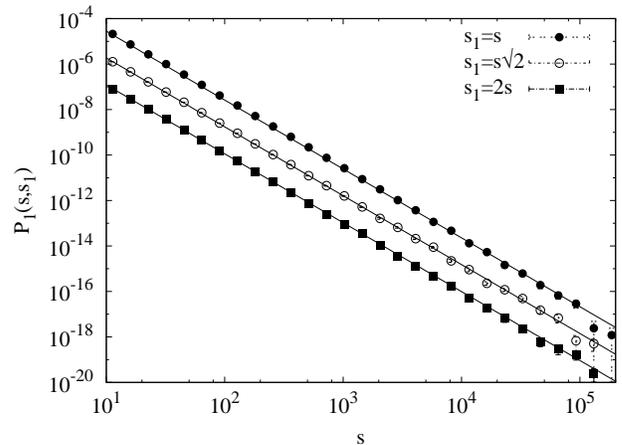}
\caption{\label{fig5}
The variation of $P_1(s,s_1)$ with $s$ is shown for the sticky model for
$s_1=s,s\sqrt{2},s$. The simulations are for $p=0.85$.
The bottom two curves have been shifted for clarity.
The solid lines are power laws with exponent $3.03$.
}
\end{figure}

For the other two models, we rely on Monte Carlo simulations. Simulations were
done for a lattice with $L=1024$ and $T=8192$. Logarithmic binning was used
with bin size $\ln(\sqrt{2})$. In the steady
state, the data was averaged over $2\times 10^8$  avalanches initiated by
toppling nearest neighbours. These multi avalanches were interspersed with
single site avalanches. Avalanches that reached the boundary were omitted from
the statistics in order to prevent strong finite size 
corrections \cite{stella1999}.

We study the variation with
$s$ of the probability distributions $P_1(s,s_1)$ and $P_2(s,s_1)$ 
with $s_1=s,s\sqrt{2}, 2 s$.
The exponents are determined through the maximum likelihood 
estimator method  
\cite{goldstein2004,newman2005}.  Let 
$P_i(s_1,s) \sim (s_1 s)^{-\tau_i/2}$, for $i=1,2$. The numerical
estimates for $\tau_1$ and
$\tau_2$ are shown in Table~\ref{table1}.
The data is shown in
Figs.~\ref{fig3} [deterministic],\ref{fig4} [stochastic], and 
\ref{fig5} [sticky], all in good agreement with Eq.~(\ref{eq:sandpilecfr0}). 
\begin{table}
\caption{\label{table1} Numerically obtained values for the exponents
$\tau_1$ and $\tau_2$ for the different models.}
 \begin{ruledtabular}
\begin{tabular}{l|c|c}
Model & $\tau_1$ & $\tau_2$ \\
\hline
Deterministic & $2.97 \pm 0.04$ & $2.98 \pm 0.04$\\
Stochastic & $3.03\pm 0.05$ & $3.00\pm 0.04$ \\
Sticky & $3.03 \pm 0.06$ & - 
\end{tabular}
\end{ruledtabular}
\end{table}

In dimensions greater than the upper critical dimension, we expect the
scaling in Eq.~(\ref{eq:sandpilecfr0}) to hold, the meanfield avalanche
exponent being $3/2$. The deviation from meanfield
should be most pronounced in two dimensions 
for which Eq.~(\ref{eq:sandpilecfr0}) has
been numerically verified. In other dimensions, we expect that an equation of
the form Eq.~(\ref{eq:sandpilecfr1}) should hold, maybe with a different
joint probability distribution. For example, in three dimensions, it will be
$P_1(s_1,s_2,s_3)$. However, the main contribution to this three point
function will be when one of the $s_i$'s is small and we retrieve an
effective two-point function.

To summarize, the two-avalanche distribution was studied  for three directed
sandpile models. While the three models have different
exponents for the single site avalanche distribution, it was shown
numerically and through a heuristic argument that the two avalanche 
distribution is the same for all three. Exact results were obtained for the
deterministic model. The robustness of the result is due to particle 
conservation layer by layer, leading to the scaling relation $\langle s
\rangle \sim t$, and is not dependent on the details of the model.

\begin{acknowledgments}
I thank O. Zaboronski,  C. Connaughton and D. Dhar for helpful discussions.
\end{acknowledgments}


\begin{thebibliography}{23}
\expandafter\ifx\csname natexlab\endcsname\relax\def\natexlab#1{#1}\fi
\expandafter\ifx\csname bibnamefont\endcsname\relax
  \def\bibnamefont#1{#1}\fi
\expandafter\ifx\csname bibfnamefont\endcsname\relax
  \def\bibfnamefont#1{#1}\fi
\expandafter\ifx\csname citenamefont\endcsname\relax
  \def\citenamefont#1{#1}\fi
\expandafter\ifx\csname url\endcsname\relax
  \def\url#1{\texttt{#1}}\fi
\expandafter\ifx\csname urlprefix\endcsname\relax\def\urlprefix{URL }\fi
\providecommand{\bibinfo}[2]{#2}
\providecommand{\eprint}[2][]{\url{#2}}

\bibitem[{\citenamefont{Bak et~al.}(1987)\citenamefont{Bak, Tang, and
  Wiesenfeld}}]{btw1987}
\bibinfo{author}{\bibfnamefont{P.}~\bibnamefont{Bak}},
  \bibinfo{author}{\bibfnamefont{C.}~\bibnamefont{Tang}}, \bibnamefont{and}
  \bibinfo{author}{\bibfnamefont{K.}~\bibnamefont{Wiesenfeld}},
  \bibinfo{journal}{Phys. Rev. Lett.} \textbf{\bibinfo{volume}{59}},
  \bibinfo{pages}{381} (\bibinfo{year}{1987}).

\bibitem[{\citenamefont{Bak et~al.}(1988)\citenamefont{Bak, Tang, and
  Wiesenfeld}}]{btw1988}
\bibinfo{author}{\bibfnamefont{P.}~\bibnamefont{Bak}},
  \bibinfo{author}{\bibfnamefont{C.}~\bibnamefont{Tang}}, \bibnamefont{and}
  \bibinfo{author}{\bibfnamefont{K.}~\bibnamefont{Wiesenfeld}},
  \bibinfo{journal}{J. Phys. A} \textbf{\bibinfo{volume}{38}},
  \bibinfo{pages}{364} (\bibinfo{year}{1988}).

\bibitem[{\citenamefont{Dhar}(2006)}]{dhar1999}
\bibinfo{author}{\bibfnamefont{D.}~\bibnamefont{Dhar}},
  \bibinfo{journal}{Physica A} \textbf{\bibinfo{volume}{369}},
  \bibinfo{pages}{29} (\bibinfo{year}{2006}).

\bibitem[{\citenamefont{Kolmogorov}(1941)}]{K41}
\bibinfo{author}{\bibfnamefont{A.~N.} \bibnamefont{Kolmogorov}},
  \bibinfo{journal}{Doklady, USSR Ac. Sci.} \textbf{\bibinfo{volume}{30}},
  \bibinfo{pages}{299} (\bibinfo{year}{1941}).

\bibitem[{\citenamefont{Kolmogorov}(1942)}]{K42}
\bibinfo{author}{\bibfnamefont{A.~N.} \bibnamefont{Kolmogorov}},
  \bibinfo{journal}{Izvestiya, USSR Ac. Sci. Phys.}
  \textbf{\bibinfo{volume}{6}}, \bibinfo{pages}{56} (\bibinfo{year}{1942}).

\bibitem[{\citenamefont{Falkovich}(2004)}]{Grisha2}
\bibinfo{author}{\bibfnamefont{G.}~\bibnamefont{Falkovich}}, in
  \emph{\bibinfo{booktitle}{Encyclopedia of Nonlinear Science}}, edited by
  \bibinfo{editor}{\bibfnamefont{A.}~\bibnamefont{Scott}}
  (\bibinfo{publisher}{Routledge}, \bibinfo{address}{New York and London},
  \bibinfo{year}{2004}).

\bibitem[{\citenamefont{Frisch}(1995)}]{frischBook}
\bibinfo{author}{\bibfnamefont{U.}~\bibnamefont{Frisch}},
  \emph{\bibinfo{title}{Turbulence: The Legacy of A. N. Kolmogorov}}
  (\bibinfo{publisher}{Cambridge University Press},
  \bibinfo{address}{Cambridge}, \bibinfo{year}{1995}).

\bibitem[{\citenamefont{Gomez et~al.}(2000)\citenamefont{Gomez, Politano, and
  Pouquet}}]{pouquet2000}
\bibinfo{author}{\bibfnamefont{T.}~\bibnamefont{Gomez}},
  \bibinfo{author}{\bibfnamefont{H.}~\bibnamefont{Politano}}, \bibnamefont{and}
  \bibinfo{author}{\bibfnamefont{A.}~\bibnamefont{Pouquet}},
  \bibinfo{journal}{Phys. Rev. E} \textbf{\bibinfo{volume}{61}},
  \bibinfo{pages}{5321} (\bibinfo{year}{2000}).

\bibitem[{\citenamefont{Falkovich et~al.}(2001)\citenamefont{Falkovich,
  Gawedzki, and Vergassola}}]{Grisha1}
\bibinfo{author}{\bibfnamefont{G.}~\bibnamefont{Falkovich}},
  \bibinfo{author}{\bibfnamefont{K.}~\bibnamefont{Gawedzki}}, \bibnamefont{and}
  \bibinfo{author}{\bibfnamefont{M.}~\bibnamefont{Vergassola}},
  \bibinfo{journal}{Rev. Modern Phys.} \textbf{\bibinfo{volume}{73}},
  \bibinfo{pages}{913} (\bibinfo{year}{2001}).

\bibitem[{\citenamefont{Connaughton
  et~al.}(2007{\natexlab{a}})\citenamefont{Connaughton, Rajesh, and
  Zaboronski}}]{CRZcfrshort}
\bibinfo{author}{\bibfnamefont{C.}~\bibnamefont{Connaughton}},
  \bibinfo{author}{\bibfnamefont{R.}~\bibnamefont{Rajesh}}, \bibnamefont{and}
  \bibinfo{author}{\bibfnamefont{O.}~\bibnamefont{Zaboronski}},
  \bibinfo{journal}{Phys. Rev. Lett} \textbf{\bibinfo{volume}{98}},
  \bibinfo{pages}{080601} (\bibinfo{year}{2007}{\natexlab{a}}).

\bibitem[{\citenamefont{Zakharov et~al.}(1992)\citenamefont{Zakharov, Lvov, and
  Falkovich}}]{zakharovBook}
\bibinfo{author}{\bibfnamefont{V.}~\bibnamefont{Zakharov}},
  \bibinfo{author}{\bibfnamefont{V.}~\bibnamefont{Lvov}}, \bibnamefont{and}
  \bibinfo{author}{\bibfnamefont{G.}~\bibnamefont{Falkovich}},
  \emph{\bibinfo{title}{Kolmogorov Spectra of Turbulence}}
  (\bibinfo{publisher}{Springer-Verlag}, \bibinfo{address}{Berlin},
  \bibinfo{year}{1992}).

\bibitem[{\citenamefont{Connaughton et~al.}(2008)\citenamefont{Connaughton,
  Rajesh, and Zaboronski}}]{CRZcfrlong}
\bibinfo{author}{\bibfnamefont{C.}~\bibnamefont{Connaughton}},
  \bibinfo{author}{\bibfnamefont{R.}~\bibnamefont{Rajesh}}, \bibnamefont{and}
  \bibinfo{author}{\bibfnamefont{O.}~\bibnamefont{Zaboronski}},
  \bibinfo{journal}{Phys. Rev. E} \textbf{\bibinfo{volume}{78}},
  \bibinfo{pages}{041403} (\bibinfo{year}{2008}).

\bibitem[{\citenamefont{Connaughton
  et~al.}(2007{\natexlab{b}})\citenamefont{Connaughton, Rajesh, and
  Zaboronski}}]{CRZinout}
\bibinfo{author}{\bibfnamefont{C.}~\bibnamefont{Connaughton}},
  \bibinfo{author}{\bibfnamefont{R.}~\bibnamefont{Rajesh}}, \bibnamefont{and}
  \bibinfo{author}{\bibfnamefont{O.}~\bibnamefont{Zaboronski}},
  \bibinfo{journal}{Physica A} \textbf{\bibinfo{volume}{384}},
  \bibinfo{pages}{108} (\bibinfo{year}{2007}{\natexlab{b}}).

\bibitem[{\citenamefont{Dhar and Ramaswamy}(1989)}]{dhar-ramaswamy1989}
\bibinfo{author}{\bibfnamefont{D.}~\bibnamefont{Dhar}} \bibnamefont{and}
  \bibinfo{author}{\bibfnamefont{R.}~\bibnamefont{Ramaswamy}},
  \bibinfo{journal}{Phys. Rev. Lett.} \textbf{\bibinfo{volume}{63}},
  \bibinfo{pages}{1659} (\bibinfo{year}{1989}).

\bibitem[{\citenamefont{Paczuski and Bassler}(2000)}]{paczuski-bassler2000}
\bibinfo{author}{\bibfnamefont{M.}~\bibnamefont{Paczuski}} \bibnamefont{and}
  \bibinfo{author}{\bibfnamefont{K.~E.} \bibnamefont{Bassler}},
  \bibinfo{journal}{Phys. Rev. E} \textbf{\bibinfo{volume}{62}},
  \bibinfo{pages}{5347} (\bibinfo{year}{2000}).

\bibitem[{\citenamefont{Kloster et~al.}(2001)\citenamefont{Kloster, Maslov, and
  Tang}}]{kloster2001}
\bibinfo{author}{\bibfnamefont{M.}~\bibnamefont{Kloster}},
  \bibinfo{author}{\bibfnamefont{S.}~\bibnamefont{Maslov}}, \bibnamefont{and}
  \bibinfo{author}{\bibfnamefont{C.}~\bibnamefont{Tang}},
  \bibinfo{journal}{Phys. Rev. E} \textbf{\bibinfo{volume}{63}},
  \bibinfo{pages}{026111} (\bibinfo{year}{2001}).

\bibitem[{\citenamefont{Tadi\'{c} and Dhar}(1997)}]{tadic-dhar1997}
\bibinfo{author}{\bibfnamefont{B.}~\bibnamefont{Tadi\'{c}}} \bibnamefont{and}
  \bibinfo{author}{\bibfnamefont{D.}~\bibnamefont{Dhar}},
  \bibinfo{journal}{Phys. Rev. Lett.} \textbf{\bibinfo{volume}{79}},
  \bibinfo{pages}{1519} (\bibinfo{year}{1997}).

\bibitem[{\citenamefont{Manna}(1991)}]{manna1991}
\bibinfo{author}{\bibfnamefont{S.~S.} \bibnamefont{Manna}},
  \bibinfo{journal}{J. Phys. A} \textbf{\bibinfo{volume}{24}},
  \bibinfo{pages}{L363} (\bibinfo{year}{1991}).

\bibitem[{\citenamefont{Mohanty and Dhar}(2002)}]{mohanty-dhar2002}
\bibinfo{author}{\bibfnamefont{P.~K.} \bibnamefont{Mohanty}} \bibnamefont{and}
  \bibinfo{author}{\bibfnamefont{D.}~\bibnamefont{Dhar}},
  \bibinfo{journal}{Phys. Rev. Lett.} \textbf{\bibinfo{volume}{89}},
  \bibinfo{pages}{104303} (\bibinfo{year}{2002}).

\bibitem[{\citenamefont{Fisher and Gelfand}(1988)}]{FG1988}
\bibinfo{author}{\bibfnamefont{M.~E.} \bibnamefont{Fisher}} \bibnamefont{and}
  \bibinfo{author}{\bibfnamefont{M.~P.} \bibnamefont{Gelfand}},
  \bibinfo{journal}{J. Stat. Phys.} \textbf{\bibinfo{volume}{53}},
  \bibinfo{pages}{175} (\bibinfo{year}{1988}).

\bibitem[{\citenamefont{Tebaldi et~al.}(1999)\citenamefont{Tebaldi, De~Menech,
  and Stella}}]{stella1999}
\bibinfo{author}{\bibfnamefont{C.}~\bibnamefont{Tebaldi}},
  \bibinfo{author}{\bibfnamefont{M.}~\bibnamefont{De~Menech}},
  \bibnamefont{and} \bibinfo{author}{\bibfnamefont{A.~L.}
  \bibnamefont{Stella}}, \bibinfo{journal}{Phys. Rev. Lett.}
  \textbf{\bibinfo{volume}{83}}, \bibinfo{pages}{3952} (\bibinfo{year}{1999}).

\bibitem[{\citenamefont{Goldstein et~al.}(2004)\citenamefont{Goldstein, Morris,
  and Yen}}]{goldstein2004}
\bibinfo{author}{\bibfnamefont{M.~L.} \bibnamefont{Goldstein}},
  \bibinfo{author}{\bibfnamefont{S.~A.} \bibnamefont{Morris}},
  \bibnamefont{and} \bibinfo{author}{\bibfnamefont{G.~G.} \bibnamefont{Yen}},
  \bibinfo{journal}{Euro. Phys. J. B} \textbf{\bibinfo{volume}{41}},
  \bibinfo{pages}{255} (\bibinfo{year}{2004}).

\bibitem[{\citenamefont{Newman}(2005)}]{newman2005}
\bibinfo{author}{\bibfnamefont{M.~E.~J.} \bibnamefont{Newman}},
  \bibinfo{journal}{Contemporary Phys.} \textbf{\bibinfo{volume}{46}},
  \bibinfo{pages}{323} (\bibinfo{year}{2005}).

\end{thebibliography}
\end{document}